\documentclass[review,number,sort&compress,3p]{elsarticle}
\usepackage{lineno}

\usepackage{graphicx}
\usepackage{epsfig,multicol,bbm}
\usepackage{amsmath}
\usepackage{appendix}

\def\lsim{\mathrel{\hbox{\rlap{\hbox{\lower4pt\hbox{$\sim$}}}\hbox{$<$}}}}

\makeatletter

\newcommand{\Rmnum}[1]{\expandafter\@slowromancap\romannumeral #1@}
\makeatother

\begin{document}

\begin{frontmatter}

\title{Channeling in solid Xe, Ar and Ne direct dark matter detectors}

\author{Nassim Bozorgnia\corref{cor1}}
\ead{nassim@physics.ucla.edu}
\address{Department of Physics and Astronomy, UCLA, 475 Portola Plaza, Los Angeles, CA 90095, USA}

\author{Graciela B. Gelmini}
\ead{gelmini@physics.ucla.edu}
\address{Department of Physics and Astronomy, UCLA, 475 Portola Plaza, Los Angeles, CA 90095, USA}

\author{Paolo Gondolo}
\ead{paolo@physics.utah.edu}
\address{Department of Physics and Astronomy, University of Utah, 115 South 1400 East \#201, Salt Lake City, UT 84112, USA\\
School of Physics, KIAS, Hoegiro 87, Seoul, 130-722, Korea}

\cortext[cor1]{Corresponding author}

\begin{abstract}
The channeling of the ion recoiling after a collision with a WIMP changes the ionization signal
in direct detection experiments, producing a larger  scintillation or ionization signal than otherwise expected. We
give estimates of the fraction of channeled recoiling ions in solid Xe, Ar and Ne  crystals
using analytic models produced since  the 1960's and 70's  to describe channeling and blocking effects.
\end{abstract}


\end{frontmatter}

\section{Introduction}

Channeling and blocking effects in crystals refer to the orientation dependence of charged ion penetration in crystals. In the ``channeling effect'' ions moving  along symmetry axes and planes in a crystal, suffer a series of small-angle scatterings  that maintain them in the open ``channels''  in between the rows or planes of lattice atoms and thus penetrate much further into the crystal than in other directions and give all their energy into electrons.  Channeled incident ions do not get close to lattice sites, where they would be deflected at large angles.  The ``blocking effect"  consists in a reduction of the flux of ions originating in lattice sites along symmetry axes and planes, due to large-angle scattering with the atoms immediately in front of the originating lattice site.

 In the context of direct dark matter detection channeling occurs when the nucleus that recoils after being hit by a WIMP (Weakly Interacting Massive Particle) moves off in a direction close to a symmetry axis or symmetry plane of the crystal. Non-channeled ions loose most of their energy into lattice atoms with which they collide. Thus channeled ions produce more scintillation and ionization than they would produce otherwise, since they give all their energy to electrons.

Ion channeling in crystals has  recently received a large amount of attention in the interpretation of direct dark matter detection data. The potential importance of this effect was first pointed out for NaI (Tl) by Drobyshevski~\cite{Drobyshevski:2007zj} and  then by the DAMA~\cite{Bernabei:2007hw} collaboration, which estimated the fraction of channeled recoils  to be very large, close to 1,  for low recoiling energies in the keV range. This estimate of the channeling effect leads to considerable shifts in the  cross section versus mass  regions of acceptable WIMP models in agreement with the DAMA/LIBRA~\cite{DAMA-bckg} annual modulation data towards lower WIMP masses. As a consequence, the comparison between the DAMA results and the null results of other experiments was affected too (see e.g. Ref.~\cite{Savage:2008er} and references therein).

In Ref.~\cite{BGG-I} we showed that  the blocking effect,  not taken into account in the DAMA evaluation, is important  and thus the fraction of recoiling  ions that are channeled is smaller than in the DAMA collaboration estimate.  The nuclei ejected from their lattice sites by WIMP collisions are initially part of a crystal row or plane, thus,  as argued originally by Lindhard~\cite{Lindhard:1965} (through what he called the ``Rule of Reversibility''), in a perfect lattice and in the absence of energy-loss processes, the probability that a particle starting from a lattice site is channeled would be zero. However, any departure of the actual lattice from a perfect lattice, e.g. due to vibrations of the lattice atoms which are always present, violates the conditions of this argument and allows for some of the recoiling lattice nuclei to be channeled. In Ref.~\cite{BGG-I} we showed that the channeling fractions for NaI (Tl) crystals would never be larger than a few percent. This implies that channeling does not change the regions in cross section versus mass of acceptable WIMP models in agreement with the DAMA/LIBRA annual modulation data at less than $7\sigma$~\cite{Savage:2010}.

Channeling in dark matter direct detection might in principle also lead to a new background free dark matter signature. This was pointed out initially in 2008 and later in 2010,  by Avignone, Creswick, and Nussinov~\cite{Avignone:2008cw} who suggested that a new type of daily modulation due to channeling could occur in NaI (and other) crystals measuring ionization or scintillation.  Such a modulation of the rate due to channeling is expected to occur at some level because the ``WIMP wind'' arrives to Earth on average from a particular direction fixed to the galaxy. Thus,  Earth's daily rotation changes the direction of the WIMP wind with respect to the crystal axes and planes,  changing the amount of recoiling ions that are channeled vs non-channeled, and therefore the amount of energy visible via scintillation or ionization. If this daily modulation could be measured, it would be a signature of dark matter without background since no other effect could depend on the orientation of the detector with respect to the WIMP wind.  Following this idea, in Ref.~\cite{BGG-DailyMod} we  provided a full computation of the effect and using our estimates of the channeling fractions in NaI obtained upper bounds to the expected amplitude of daily modulation due to channeling in NaI crystals.  We found large daily modulation amplitudes of the signal rate, even of the order of 10\% for some WIMP candidates. However, even the largest amplitudes would not be observable in the 13 years of combined DAMA/NaI and DAMA/LIBRA~\cite{DAMA-bckg} data because of their large background. In the hope that the daily modulation due to channeling could be observable in future dark matter experiments, we have provided estimates of the channeling fractions of recoiling ions also in Ge, Si~\cite{BGG-II}, and CsI~\cite{BGG-III} crystals, and in the present paper we study noble gas crystals.

The level of background in future experiments is a crucial element to determine if the  daily modulation is observable and the advantage of using solidified noble gas detectors is the possibility of achieving a low background level for WIMP searches~\cite{Balakishiyeva-2010}. Xenon and Neon have no long life radioisotopes and thus contain no intrinsic background source of radiation. Argon has the drawback that it contains $^{39}\textrm{Ar}$ beta source which induces electron recoil signature in the detector. Solid or crystallized Xenon detector can be ideal for dark matter searches, whereas a crystallized Xenon detector would be necessary to have channeling. At present, solid Xe crystals are used by the Solid Xenon R\&D Project~\cite{Yoo-2010}.

In this paper we compute the geometric channeling fractions of recoiling ions in solid Xe, Ar, and Ne crystals. Here ``geometric channeling fraction'' refers to assuming that the distribution of recoil directions is isotropic. In reality, in a dark matter direct detection experiment, the distribution of recoil directions is expected to be peaked in the direction of the average WIMP flow. At room temperature and pressure  Xe, Ar and Ne are noble gases. At temperatures below 161.45 K, 83.80 K and 24.56 K respectively (at room pressure) they become solids. All of them form   monatomic  face-centered cubic (f.c.c.) crystals (see Appendix A).

Our calculation  is based on  the classical analytic models developed in the 1960's and 70's, in particular by Lindhard and collaborators~\cite{Lindhard:1965, Dearnaley:1973, Andersen:1967, Morgan-VanVliet, VanVliet, Andersen-Feldman, Komaki:1970,  Appleton-Foti:1977, Hobler}. In these models the discrete series of binary collisions of the propagating ion with atoms is replaced by interactions between the ion and uniformly charged strings or planes. The   screened atomic Thomas-Fermi potential is averaged over a direction parallel to a row or a plane of lattice atoms. This averaged potential is considered to be uniformly smeared along the row or plane of atoms, respectively, which is a good approximation if the propagating ion interacts with many lattice atoms in the row or plane by a correlated series of many consecutive glancing collisions with lattice atoms. We consider just one row or one plane, which simplifies the calculations and is correct except at the lowest energies we consider.

We proceed  in a similar manner as we did for Si and Ge in Ref.~\cite{BGG-II}, and we refer to that paper for formulas and other details. There are several good analytic approximations of the screened potential. As in Ref~\cite{BGG-II}, here we use  Moli\`{e}re's approximation for the atomic potential, following the work of Hobler~\cite{Hobler} and Morgan and Van Vliet~\cite{Morgan-VanVliet, VanVliet}. In  Moli\`{e}re's approximation~\cite{Gemmell:1974ub} the axial continuum potential, as a function of the transverse distance $r$ to the string, is
\begin{linenomath}
 \begin{equation}
 U(r)=\left(2Z_1 Z_2 e^2/d\right)f(r/a)
 =E\psi_1^2f(r/a),
 \end{equation}
 \end{linenomath}
 where $E$ is the energy of the propagating particle and  $\psi_{1}^2=2Z_{1}Z_{2}e^2/(E d)$. Here $Z_1$, $Z_2$ are the atomic numbers of the recoiling and lattice nuclei respectively, $d$ is the spacing between atoms in the row, $a$ is the Thomas-Fermi screening distance, $a= 0.4685 {\text {\AA} } (Z_1^{1/2} + Z_2^{1/2})^{-2/3} $~\cite{Barrett:1971, Gemmell:1974ub} and   $E= Mv^2/2$ is the kinetic energy of the propagating ion. Moli\`{e}re's screening function~\cite{Gemmell:1974ub} for the continuum potential is $f(\xi)=\sum_{i=1}^{3}{\alpha_i K_0(\beta_i \xi)}$. Here $K_0$ is the zero-order modified Bessel function of the second kind, and the dimensionless  coefficients  $\alpha_i$ and  $\beta_i$ are  $\alpha_i=\{ 0.1, 0.55, 0.35 \}$  and $\beta_i=\{ 6.0, 1.2, 0.3 \}$ ~\cite{Watson:1958}, for $i=1,2,3$. The string of crystal atoms is at $r=0$. In our case, $E$ is the recoil energy imparted to the ion in a collision with a WIMP,
 \begin{linenomath}
\begin{equation}
E = \frac{|\vec{\bf q}|^2}{2M},
\end{equation}
\end{linenomath}
and $\vec{\bf q}$ is the recoil momentum.

\begin{figure}
\begin{center}
  \includegraphics[height=135pt]{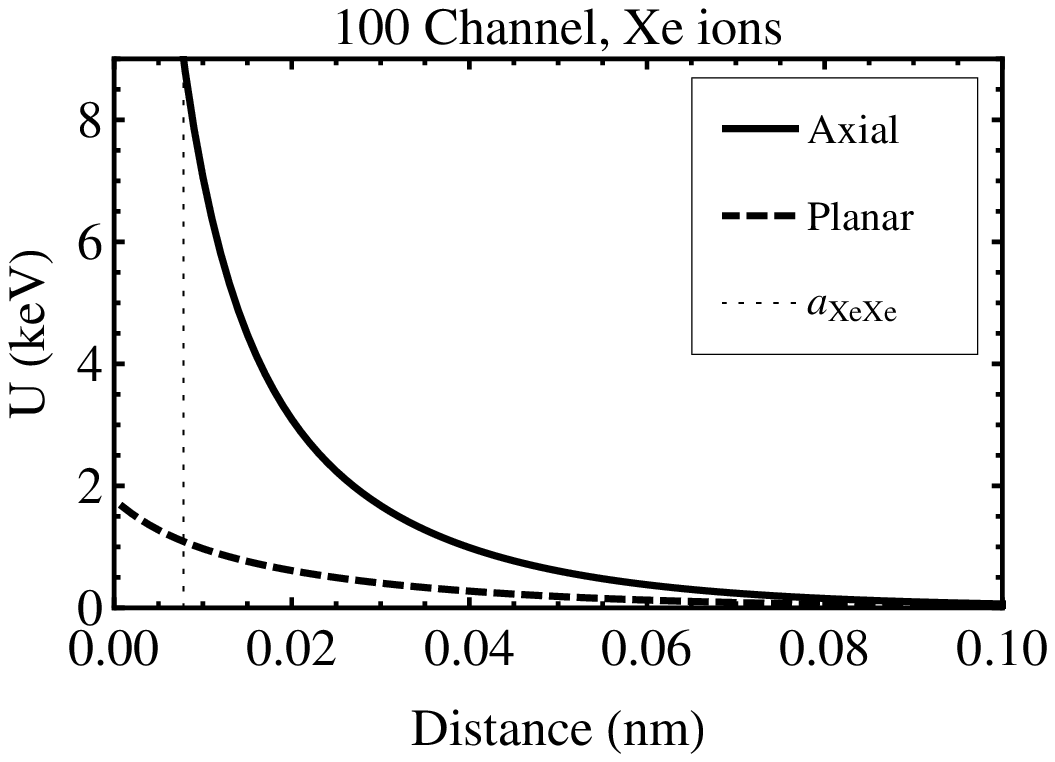}
  \includegraphics[height=135pt]{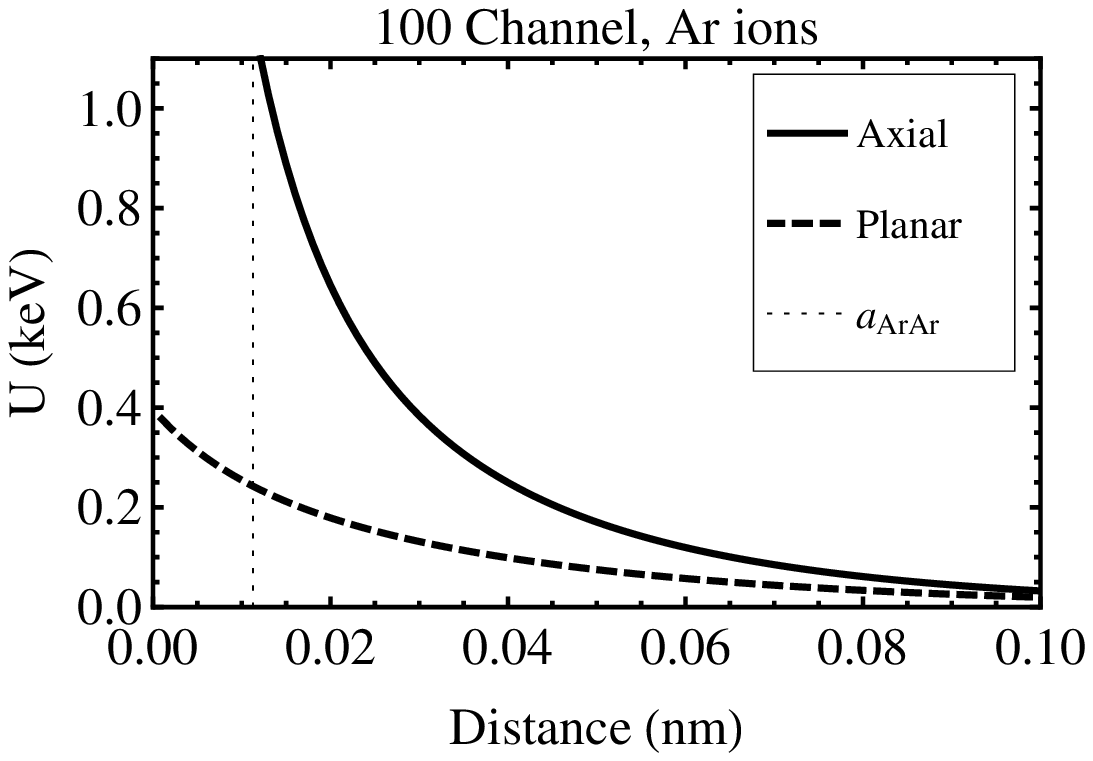}\\
  \includegraphics[height=137pt]{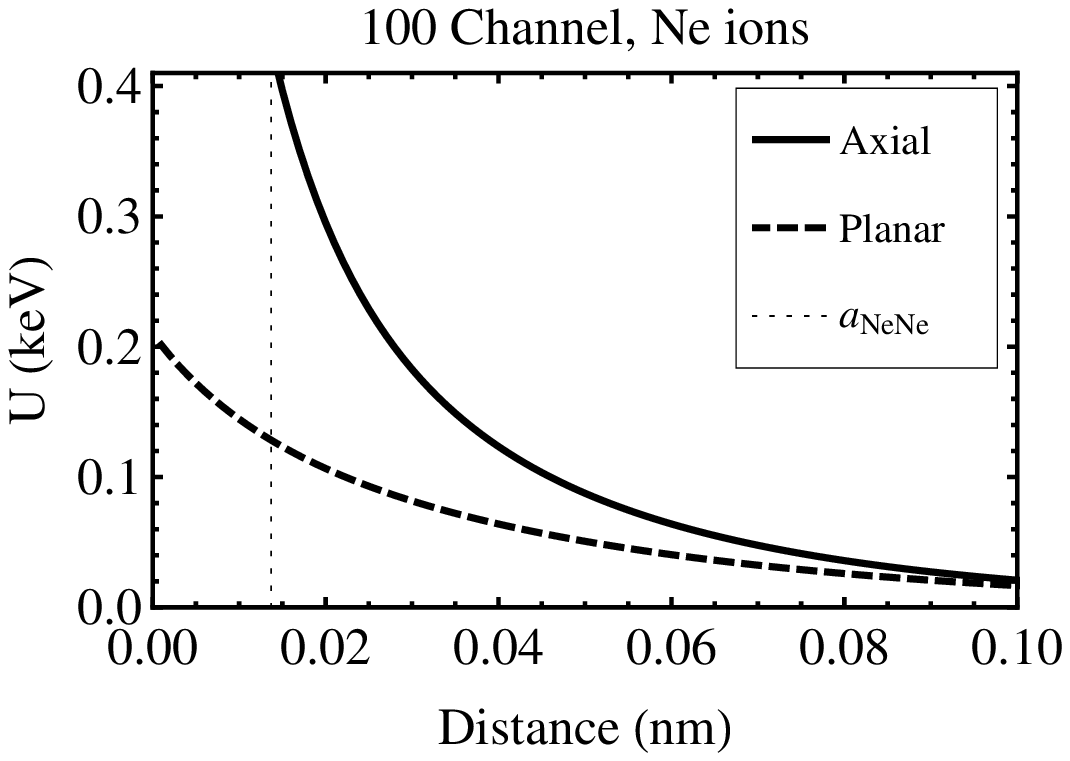}\\
  \caption{Continuum axial (solid lines) and planar (dashed lines) potentials as function of the distance from the row or plane of lattice atoms, respectively, for (a) Xe ions, (b) Ar ions, and (c) Ne ions, propagating  in the $<$100$>$ axial and \{100\} planar channels of a Xe, Ar, and Ne crystal respectively. The screening radius shown as a vertical line is $a_{\rm XeXe}=0.007808$ nm for Xe, $a_{\rm ArAr}=0.01126$ nm for Ar and $a_{\rm NeNe}=0.01370$ nm for Ne (see App. A).}
  \label{U}
\end{center}
\end{figure}

The continuum planar potential  in Moli\`{e}re's approximation~\cite{Gemmell:1974ub}, as a function of the distance $x$ perpendicular to the plane, is
\begin{linenomath}
 \begin{equation}
 U_p(x)=\left(2\pi n Z_1 Z_2 e^2 a\right)f_p(x/a)
 =E\psi_a^2f_p(x/a),
 \end{equation}
 \end{linenomath}
where $\psi_a^2=2\pi n Z_1 Z_2 e^2 a/E$. Here $n= N d_{pch}$ is the average number of atoms per unit area, where $N$ is the atomic density and $d_{pch}$ is the width of the planar channel, i.e.  the  interplanar spacing (thus, the average distance of atoms within a plane is $d_p=1/ \sqrt{N d_{pch}}$). The subscript p denotes ``planar" and $f_p(\xi)=\sum_{i=1}^{3}{(\alpha_i/\beta_i) \exp(-\beta_i \xi)}$, where the coefficients $\alpha_i$ and $\beta_i $ are the same as above. The plane is at $x=0$. Also, the axial channel width which we call $d_{\rm ach}$ is defined in terms of the interatomic distance $d$ in the corresponding row as $d_{\rm ach} = 1/\sqrt{Nd}$. Examples of axial and planar continuum potentials, generically called  $U$, for Xe, Ar, and Ne ions propagating in the $<$100$>$ axial and \{100\} planar channels of a Xe, Ar, and Ne crystal respectively are shown in Fig.~\ref{U}.

Lindhard proved that for channeled particles the longitudinal component of the velocity, i.e. the component along the direction of the row or plane of the velocity, may be treated as constant (if energy loss processes are neglected). Then, in the continuum model, the trajectory of the ions can be completely described in terms of the transverse direction, perpendicular to the row or plane considered. For small  angle $\phi$  between the ion's trajectory and  the atomic  row (or plane) in the direction perpendicular to  the row (or plane), the  so called ``transverse energy"
\begin{linenomath}
\begin{equation}
E_{\perp} = E \sin^2\phi + U
\label{E-perp}
\end{equation}
\end{linenomath}
is conserved. In Eq. \ref{E-perp} relativistic corrections are neglected.

The conservation of the transverse energy provides a definition of the minimum distance of approach to the string, $r_{\rm min}$ (or to the plane of atoms $x_{\rm min}$), at which the trajectory of the ion makes  a zero angle with the string (or plane), and also of the angle $\psi$ at which the ion exits from the string (or plane), i.e. far away from it where $U \simeq0$. In reality the furthest position from   a string or plane of atoms is  the middle of the channel (whose width we call  $d_{ach}$ for an axial channel and $d_{pch}$ for a planar channel).  Thus, for an axial channel
\begin{linenomath}
\begin{equation}
\label{eq:consetrans}
E_{\perp}= U(r_{\rm min}) =  E \psi^2 +U(d_{ach}/2).
\end{equation}
\end{linenomath}

The continuum model implies that the net deflection due to the succession of impulses from the peaks of the potential is identical to the  deflection due to a force $-U'$. This is only so if the ion never approaches so closely  any individual atom that it suffers a large-angle collision.  Lindhard  proved that for a string of atoms this is so only if $U''(r) < 8E/d^2$, where the double prime denotes the second derivative with respect to $r$. Replacing the inequality in this equation by an equality defines an energy dependent critical distance $r_c$ such that  $r >  r_c$ for the continuum model to be valid. Morgan and Van Vliet~\cite{Morgan-VanVliet} use 5 instead of 8 in this equation, because this agrees better with their simulations of channeling in copper crystals. As in Ref.~\cite{BGG-II} and following Hobler~\cite{Hobler} (who studied channeling in Si crystals to be able to avoid it in ion-implantation) we use Morgan and Van Vliet's equation to define $r_c$, i.e.
\begin{linenomath}
 \begin{equation}
U''(r_c) =  \frac{5}{d^2} E.
\label{MVU''}
\end{equation}
\end{linenomath}
For a ``static lattice'', that here means a perfect lattice in which all vibrations are neglected, the critical distance of approach $r_c$ is given in Eq. 2.15 of Ref.~\cite{BGG-II}. We use  an approximate analytic expression for $r_c$ obtained by fitting a degree nine polynomial in  the parameter $\sqrt{\alpha}= (Z_1 Z_2 e^2 d/ a^2 E)^{1/2}$ to the exact solution of Eq.~\ref{MVU''}. The expression for $r_c$ obtained in this way (Eq. 2.15 of Ref.~\cite{BGG-II}) is valid for recoil energies $E >$ 3 keV for Xe, and for $E>1$ keV for Ar and Ne.

Since $r_c$ is the smallest possible minimum distance of approach to the string of a channeled propagating ion  for a given energy $E$, i.e. $r_{\rm min} > r_c$, and  the potential $U(r)$  decreases monotonically with increasing $r$, then $U(r_{\rm min}) < U(r_c)$. Using Eq.~\ref{eq:consetrans}, this condition translates into an upper bound on $\psi$, $\psi < \psi_{c}(E)$ where $\psi_{c}(E)$ is the critical channeling  angle for the particular axial channel, i.e. the maximum angle  the  propagating ion can make with the string far  away from it (in the middle of the channel) if the ion is channeled. We proceed similarly for planar channels to get the critical angles $\psi^p_c(E)$. However, the treatment of planar channels presents some complications.

The breakdown of the continuum theory for a planar channel is more involved than for an axial channel because  the atoms in the plane contributing to the scattering of the propagating ion are usually displaced laterally within the plane with respect to the ion's trajectory. Thus the moving ion does not encounter atoms at a fixed separation or at fixed impact parameter as is the case for a row. To find the  static critical distance $x_c$ of a planar channel we follow the procedure of defining a  ``fictitious string'', introduced by Morgan and Van Vliet~\cite{Morgan-VanVliet, Hobler}. They reduced the problem of scattering from a plane of atoms to the scattering from an equivalent fictitious row contained in a strip of width $2R$ centered on the projection of the ion path onto the plane of atoms.  Along the fictitious row, the characteristic distance between atoms needs to be estimated using data or simulations which are not available for Xe, Ar and Ne crystals. As explained in Ref.~\cite{BGG-II}, this characteristic distance  depends on the width $2R$ of the strip considered. For $R$, Morgan and Van Vliet used the impact parameter in an ion-atom collision corresponding to a deflection angle of the order of ``the break-through" angle $\sqrt{U_p(0)/E}$. This is the minimum angle at which an ion of energy $E$ must approach the plane from far away (so that the initial potential can be neglected) to overcome the potential barrier at the center of the plane at $x=0$ (namely, so that $E_\perp= U_p(0)$). For small scattering angles, the deflection angle $\delta$ is related to the impact parameter, in this case $R$, as (see e.g. Eq. $2.1'$ of Lindhard~\cite{Lindhard:1965})
 \begin{linenomath}
\begin{equation}
2 E \delta= -d~U'(R),
\label{U'}
\end{equation}
\end{linenomath}
where $U'$ is the derivative of the axial continuum potential, and Morgan and Van Vliet define $R$ by taking $\delta=\sqrt{U_p(0)/E}$.

As described in Ref.~\cite{BGG-II}, we decided to keep  the Morgan and Van Vliet definition for $R$, because Hobler~\cite{Hobler} found that it is in quite good agreement with the binary collision simulations and data of B and P ions propagating in Si for energies of about 1 keV and above. We used an approximate analytical solution for $R$ (Eq. 2.22 of Ref.~\cite{BGG-II}) obtained by fitting a degree five polynomial in $(\ln{y})$ where $y=Z_1 Z_2 e^2/a \sqrt{E U_p(0)}$, to the exact numerical solution of Eq.~\ref{U'}. The planar critical distance $x_c$ we obtained (Eq. 2.23 of Ref.~\cite{BGG-II}) is  valid for $E < 7$ GeV for Xe, $E < 160$ MeV for Ar, and $E< 20$ MeV for Ne. These conditions and those after Eq.~\ref{MVU''} provide the energy ranges for which our channeling fraction estimates are valid. Within its range of validity, the percentage error of the analytic approximation we used for $x_c$ is less than 9\%.

\begin{figure}
\begin{center}
  \includegraphics[height=210pt]{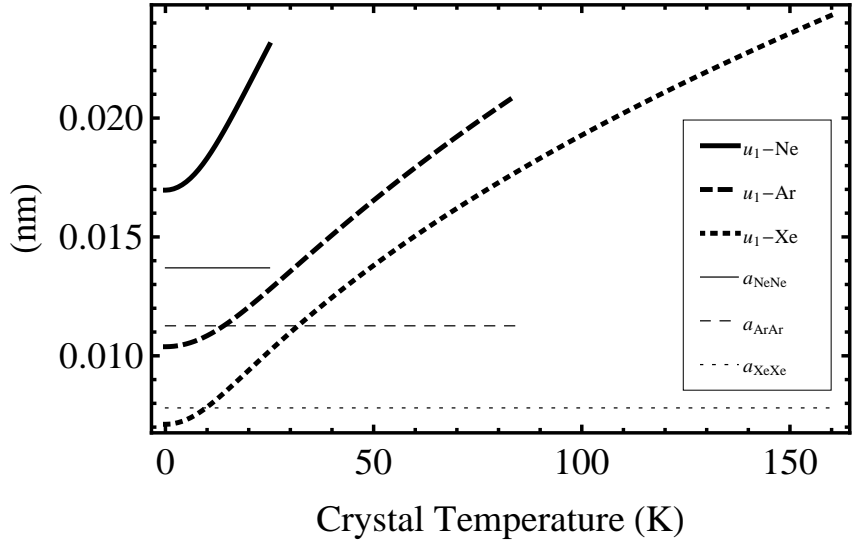}
  \caption{Plot of $u_1(T)$ for Ne (thick solid line), Ar (thick dashed line), and Xe (thick dotted line) as a function of the crystal temperature. The plots are cut at the respective melting temperatures.}
  \label{figu1}
\end{center}
\end{figure}

The critical distances of approach in a non-static lattice depend on the temperature, through the vibration of the atoms in the lattice (thermal expansion effects are negligible, as shown in Appendix  B of Ref.~\cite{BGG-III}). We use the Debye model to account  for the vibrations of the atoms in a crystal.  The one dimensional rms vibration amplitude $u_1$ of the atoms in a crystal in this model (see Eqs. 2.27 and 2.28 in  Ref.~\cite{BGG-II}) is plotted in Fig.~\ref{figu1} for Xe, Ar, and Ne crystals as function of the temperature $T$ up to their respective melting points. The crystals in the Solid Xenon R\&D Project experiment will be operating at 77.2 K and higher~\cite{Yoo-Private}.

\begin{figure}
\begin{center}
  \includegraphics[height=150pt]{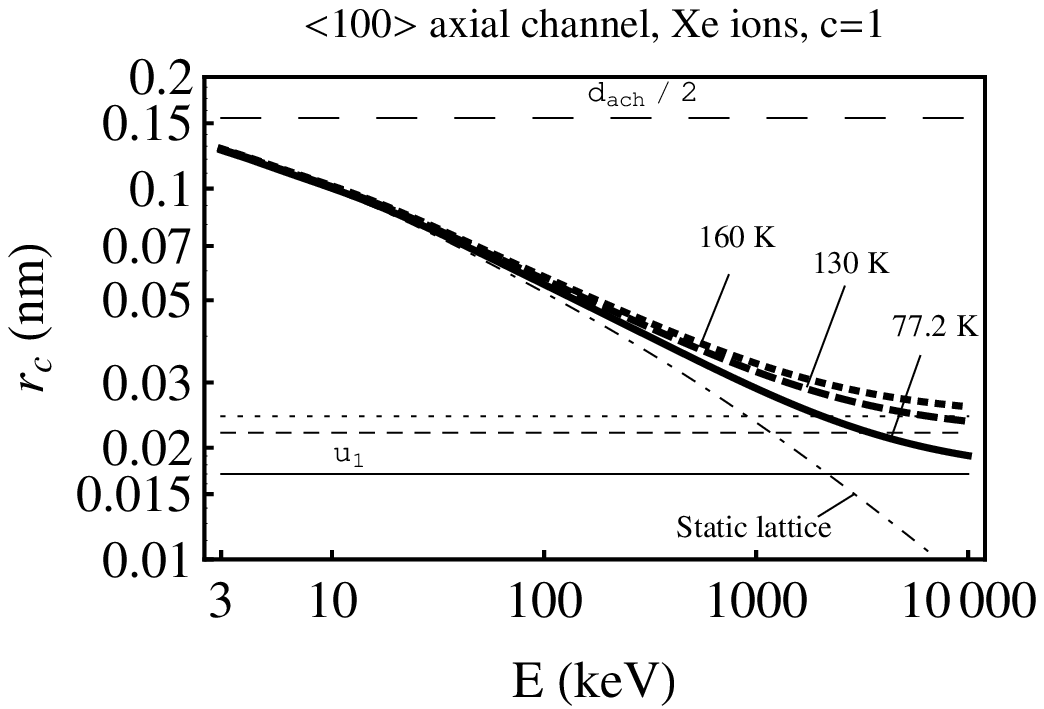}
  \includegraphics[height=150pt]{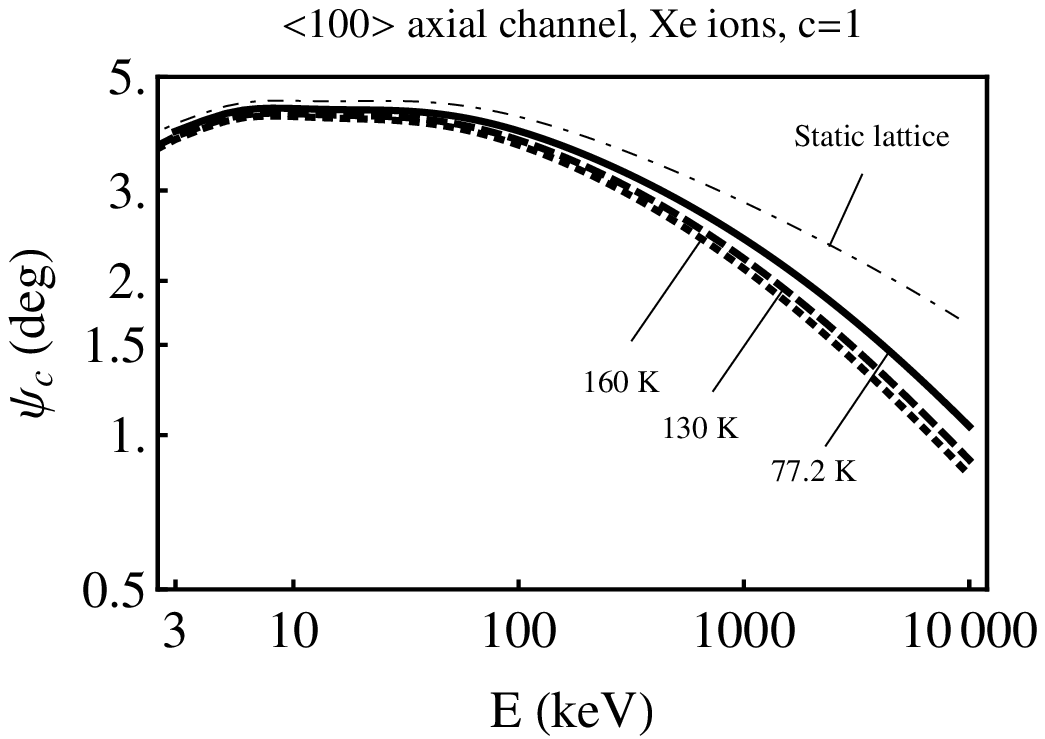}\\
  \caption{Static (thin dot-dashed line) and temperature corrected with $c_1=c_2=c=1$ (thick lines) (a) critical distances of approach (and $u_1(T)$) and (b) the corresponding critical channeling angles, for $T=77.2$ K, $T=130$ K, and $T=160$ K as a function of the energy of propagating Xe ions in the $<$100$>$ axial channel of a Xe crystal.}
  \label{rc-Xe100-DiffT-c1}
\end{center}
\end{figure}

\begin{figure}
\begin{center}
  \includegraphics[height=150pt]{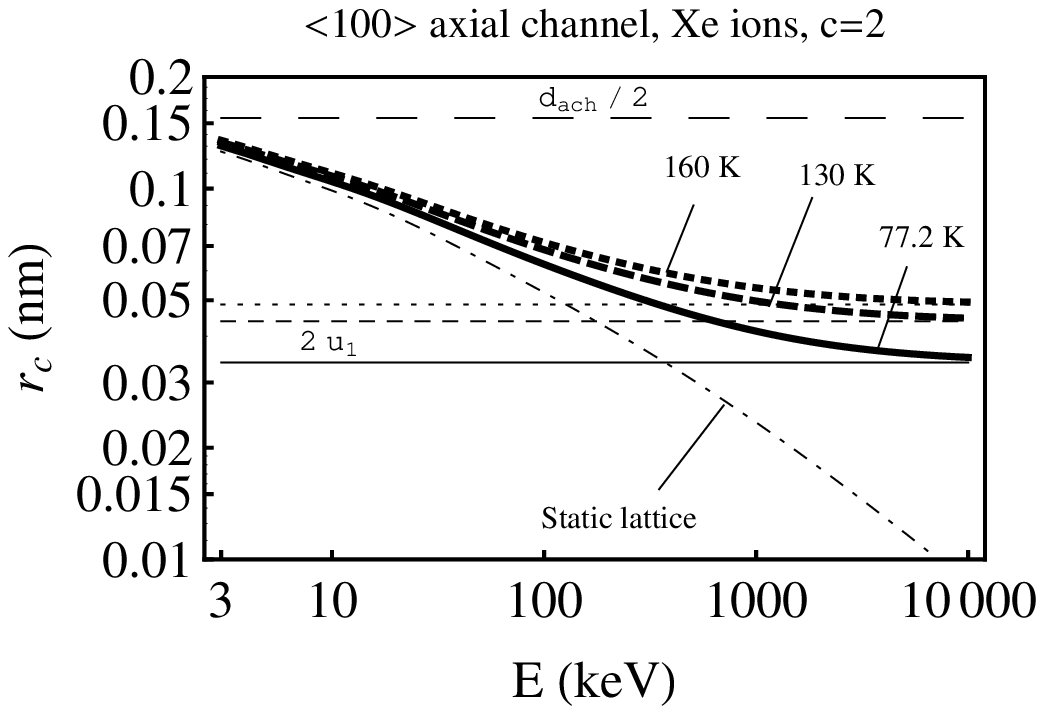}
  \includegraphics[height=150pt]{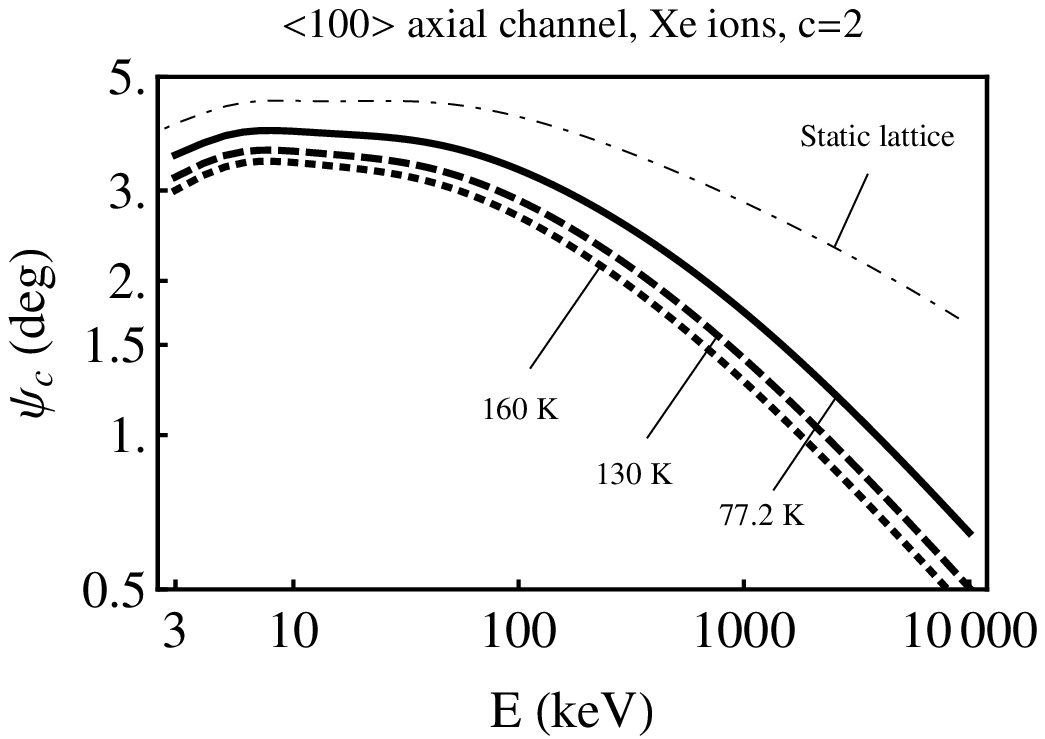}\\
  \caption{Same as Fig.~\ref{rc-Xe100-DiffT-c1} but with $c_1=c_2=c=2$.}
  \label{rc-Xe100-DiffT-c2}
\end{center}
\end{figure}

\begin{figure}
\begin{center}
  \includegraphics[height=150pt]{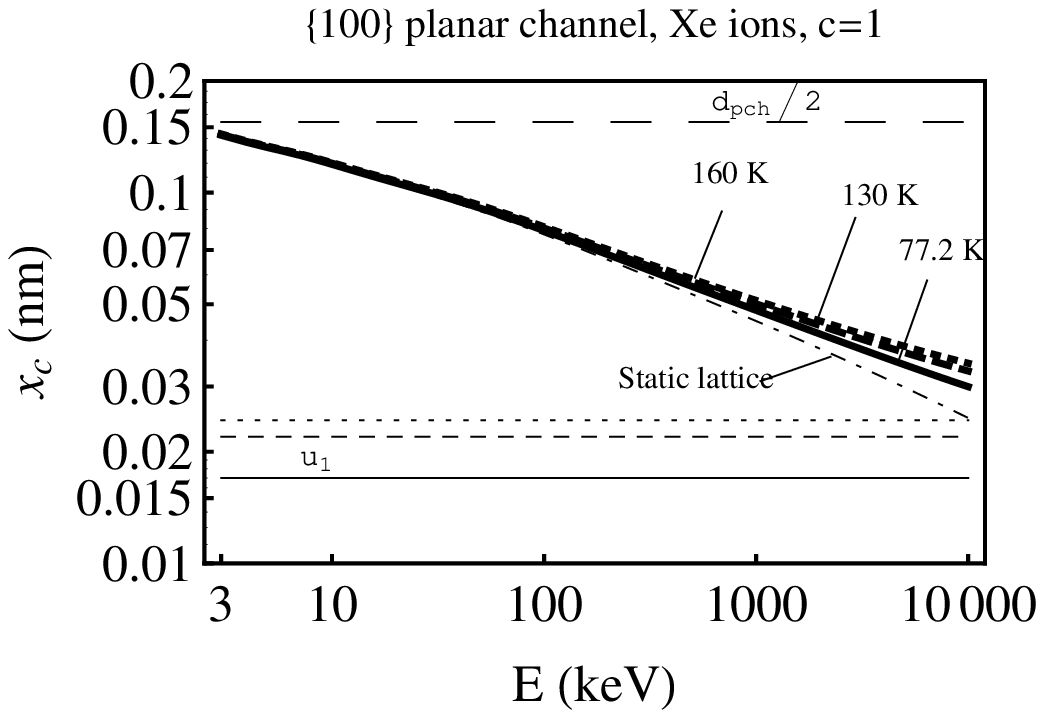}
  \includegraphics[height=150pt]{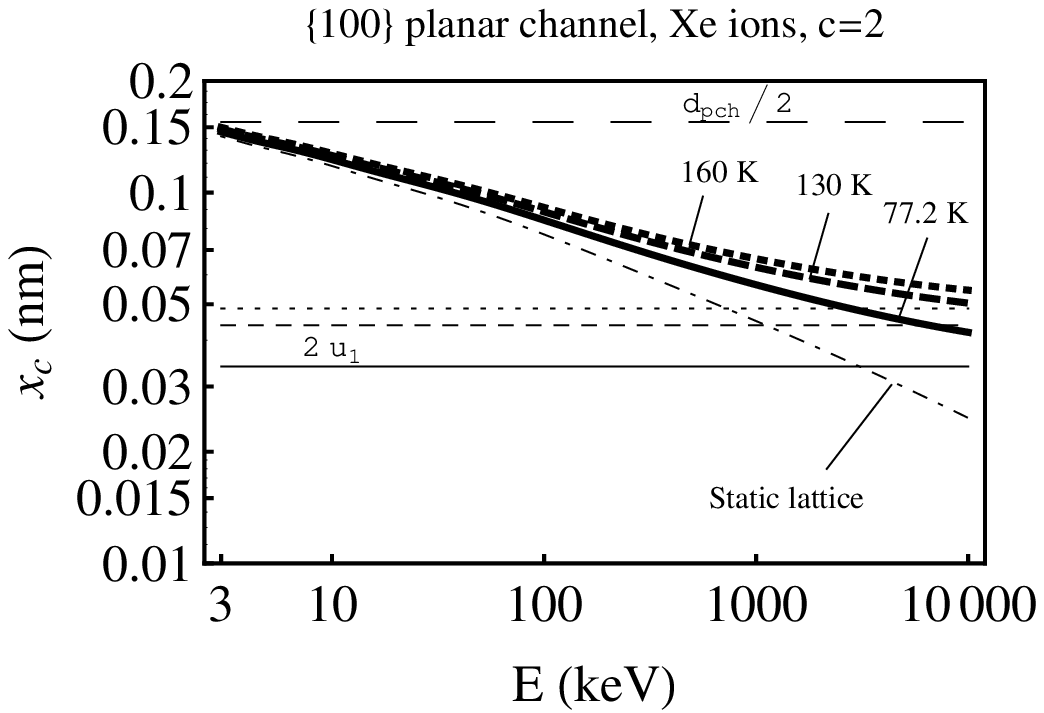}\\
  \caption{Static (thin dot-dashed line) and temperature corrected (thick lines) with (a) $c_1=c_2=c=1$ and (b) $c_1=c_2=c=2$ critical distances of approach for $T=77.2$ K, $T=130$ K, and $T=160$ K as a function of the energy of propagating Xe ions in the \{100\} planar channel of a Xe crystal.}
  \label{xc-Xe100-DiffT}
\end{center}
\end{figure}

In principle there are modifications to the continuum potentials due to thermal effects, but we take into account thermal effects  in the crystal through a modification of the critical distances found originally by Morgan and Van Vliet~\cite{Morgan-VanVliet}  and later by Hobler~\cite{Hobler} to provide good agreement with simulations and data. For axial channels it consists of taking the temperature corrected critical distance $r_c(T)$  to be $r_c(T)= \sqrt{r^2_c(E) + [c_1 u_1(T)]^2}$. For planar channels the situation is more complicated, because some references give a linear and others a quadratic relation between $x_c(T)$ and $u_1$. Following Hobler~\cite{Hobler} we use an equation similar to that for axial channels, namely $x_c(T)= \sqrt{x^2_c(E) + [c_2 u_1(T)]^2}$. The dimensionless factors $c_1$ and $c_2$ were found to be numbers between 1 and 2 in the prior literature, for different crystals and propagating ions (see Ref.~\cite{BGG-I} and~\cite{BGG-II}). Using the temperature corrected critical distances of approach  $r_c(T)$ and $x_c(T)$  or  the static lattice critical distances $r_c$ and $x_c$ (Eqs. 2.15 and 2.23 of Ref.~\cite{BGG-II}), we obtain the corresponding temperature corrected critical axial and planar channeling angles, $\psi_c$ and  $\psi^p_c$ respectively. As $c_1$ and $c_2$ increase, $r_c(T)$ and $x_c(T)$ increase and the corresponding critical angles decrease, respectively, at a given temperature. Taking $c_1=c_2=0$ corresponds to considering only a static lattice,  with no temperature effects included. Examples of critical distances and angles are shown in  Figs.~\ref{rc-Xe100-DiffT-c1} to \ref{xc-Xe100-DiffT}, for $c_1=c_2=c$ and $c=1$ or  $c=2$ as indicated.
Fig.~\ref{rc-Xe100-DiffT-c1} clearly shows how the critical distances and angles change with temperature for Xe ions propagating in the  $<$100$>$ axial channel of a Xe crystal, with temperature effects computed with $c_1=c_2=1$. At small energies the static critical distance of approach is much larger than the vibration amplitude $u_1$, so temperature corrections are not important. As the energy increases, the static critical distance of approach decreases, and when it becomes negligible with respect to the vibration amplitude $u_1$, the temperature corrected critical distance $r_c$ becomes equal  to $c_1 u_1$. Fig.~\ref{rc-Xe100-DiffT-c2} shows the same as Fig.~\ref{rc-Xe100-DiffT-c1} but for $c_1=c_2=2$.
Figs.~\ref{xc-Xe100-DiffT}(a) and \ref{xc-Xe100-DiffT}(b) show the critical distances at several temperatures for Xe ions  propagating in the \{100\} planar channel with $c_1=c_2=1$ and $c_1=c_2=2$ respectively.

\section{Channeling fractions}

In  our model, a recoiling ion is channeled if the collision ion-WIMP happens at a distance large enough from  the string  or plane to which the ion belongs. Namely, channeling happens if the initial position of the recoiling motion is $r_i > r_{i,\rm min}$ or $x_i> x_{i,\rm min}$ for an axial or planar channel respectively. We define $E_{\perp}$ in terms of the initial recoil energy $E$ of the propagating ion, the angle of the initial recoil momentum  with respect to the particular string or plane of atoms $\phi_i$, and the initial position $r_i$. For an axial channel, the condition
\begin{linenomath}
\begin{equation}
E_{\perp}(E,\phi_i,r_i)= U(r_{\rm min}) < U(r_c),
\label{ChanCond-CA}
\end{equation}
\end{linenomath}
i.e. $E_{\perp}(E,\phi_i,r_{i,\rm min})= U(r_c(E))$ defines the distance $r_{i,\rm min}$ in terms of $r_c$. The equivalent formulas for axial channels define $x_{i,\rm min}$ in terms of $x_c$. In Ref.~\cite{BGG-I}, we derived an analytic expression  for $r_{i,\rm min}$ and $x_{i,\rm min}$ using Lindhard's approximation to the potential (see Eqs. 5.9 and 5.11 of Ref.~\cite{BGG-I}).

We take the initial distance distribution (in $r$ or $x$ respectively) of the colliding atom to be a Gaussian with  a  one dimensional dispersion $u_1$, and we obtain the probability of channeling for each individual channel  by integrating the Gaussian between the minimum initial distance and infinity (a good approximation to the radius of the channel; see Ref.~\cite{BGG-I} for details). The dependence of these probabilities on the critical distances enter in the argument of an exponential or an erfc function. Thus any uncertainty in our modeling of the critical distances becomes exponentially enhanced in the channeling fraction.  This is the major difficulty of the analytical approach we are following.

In order to obtain the total geometric channeling fraction we need to sum over all the individual channels we consider. The integral over initial directions is computed using HEALPix~\cite{HEALPix:2005} (see Appendix B of Ref.~\cite{BGG-I}). Taking only the channels with lowest crystallographic indices, 100, 110 and 111, we have a total of 26 axial and planar channels (see Appendix A). We treat channeling along different channels as independent events when computing the probability that an ion enters any of the available channels. In Appendix D of Ref.~\cite{BGG-I} we showed that we can obtain an upper limit to the channeling probability of overlapping channels by replacing the intersection of the complements of the integration regions in Eqs. 5.2 and 5.4 of Ref.~\cite{BGG-I} with the inscribed cylinder of radius $r_{\rm MIN}$ equal to the minimum of the $r_{i,\rm min}$ or $x_{i,\rm min}$ among the overlapping channels. We find that this method gives results practically indistinguishable from those obtained assuming that channeling along different channels are independent events.

\begin{figure}
\begin{center}
  \includegraphics[height=170pt]{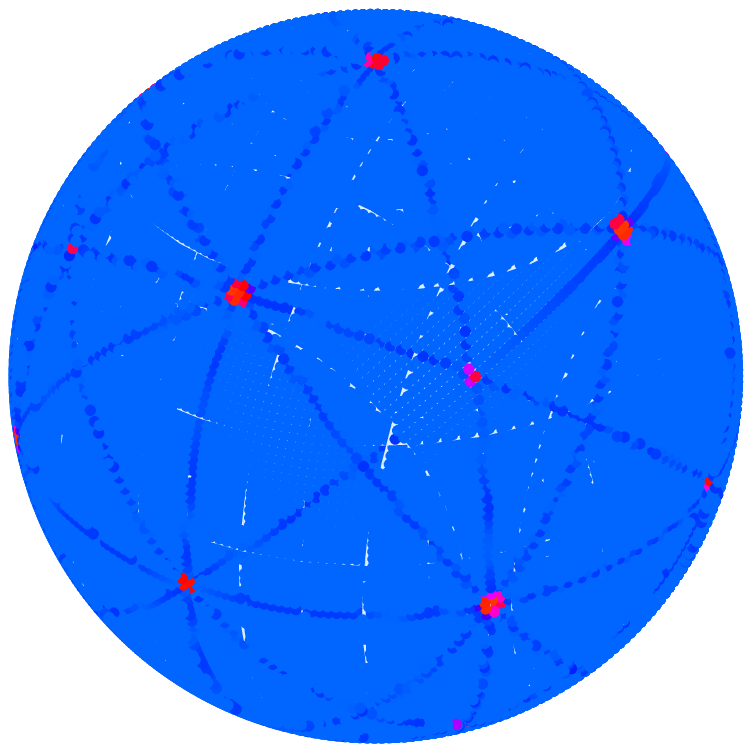}
  \includegraphics[height=170pt]{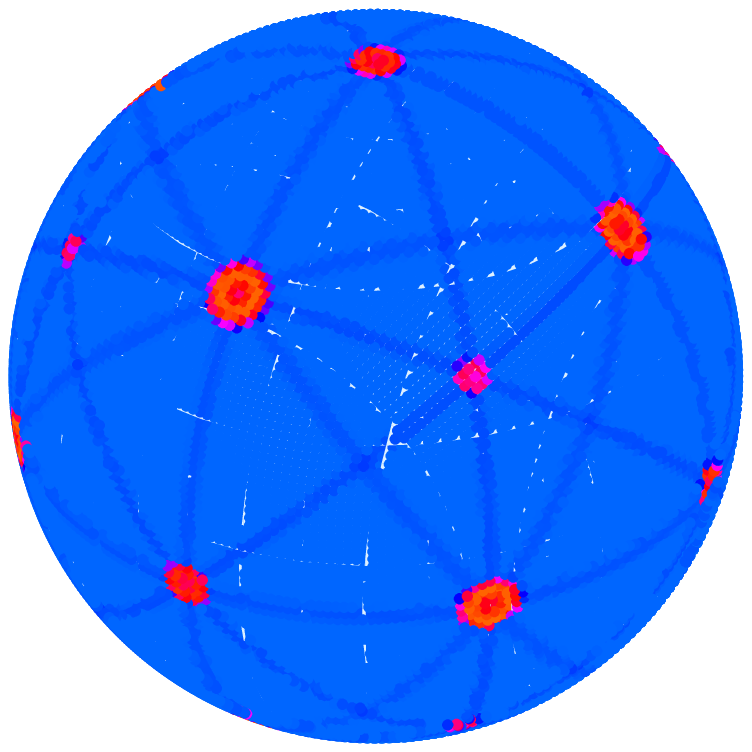}
  \includegraphics[height=25pt]{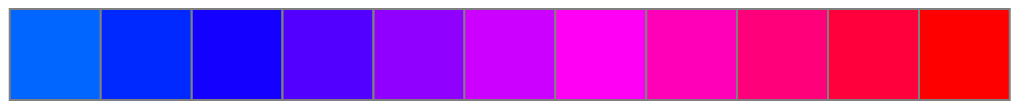}\\
  \caption{(color online) Geometric channeling probabilities as function of the initial recoil direction for (a) a 1 MeV Xe ion propagating in a Xe crystal  at 160 K and (b) a  20 keV Ne ion propagating in a Ne crystal   at 23 K (with $c_1=c_2=1$). The probability is computed for each direction and plotted on a sphere using the HEALPix pixelization. The colors/grayscale shown in the horizontal bar between black and white (blue and red online) corresponds to values between 0 and 1 in increments of 0.1.}
  \label{Our-HEALPIX}
\end{center}
\end{figure}

Fig.~\ref{Our-HEALPIX} shows the channeling probability  computed for each initial recoil direction $\hat{\bf q}$ plotted on a sphere using the HEALPix pixelization for  (a) a 1 MeV Xe ion propagating in a Xe crystal at 160 K and (b) a 20 keV Ne ion propagating in a Ne crystal at 23 K ($c_1=c_2=1$ is assumed for the temperature effects). The white and black (red and blue online) indicate a channeling probability of 1 and zero, respectively (see the colors/grayscale in the figure).

\begin{figure}[t]
\begin{center}
  \includegraphics[height=145pt]{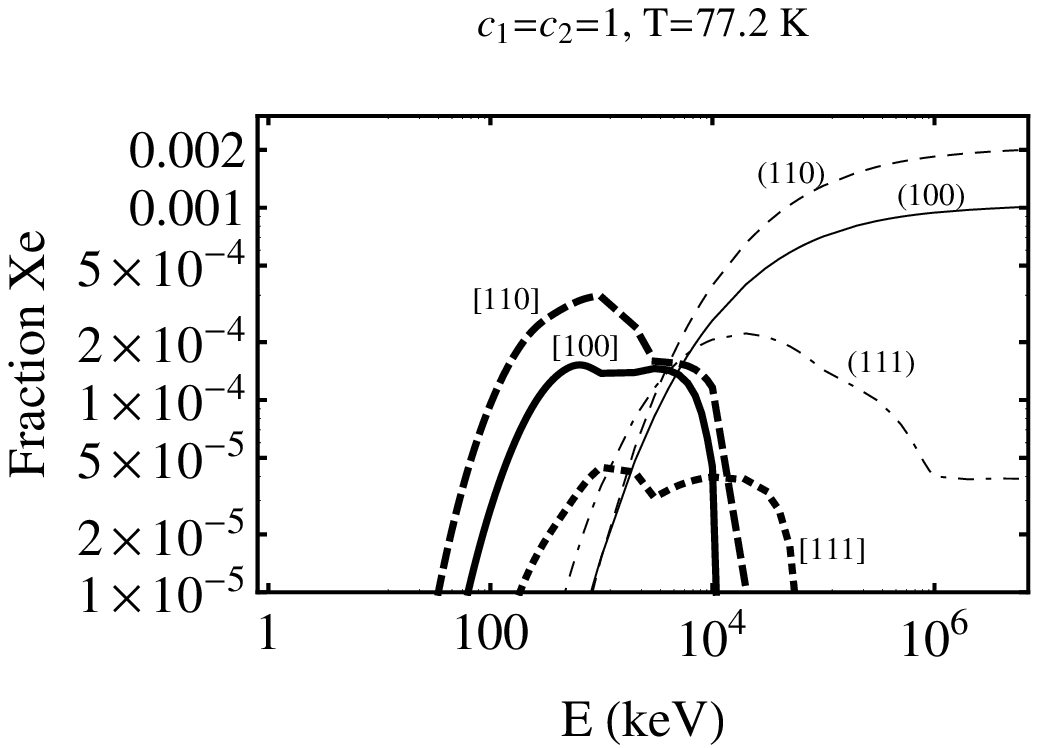}
  \includegraphics[height=145pt]{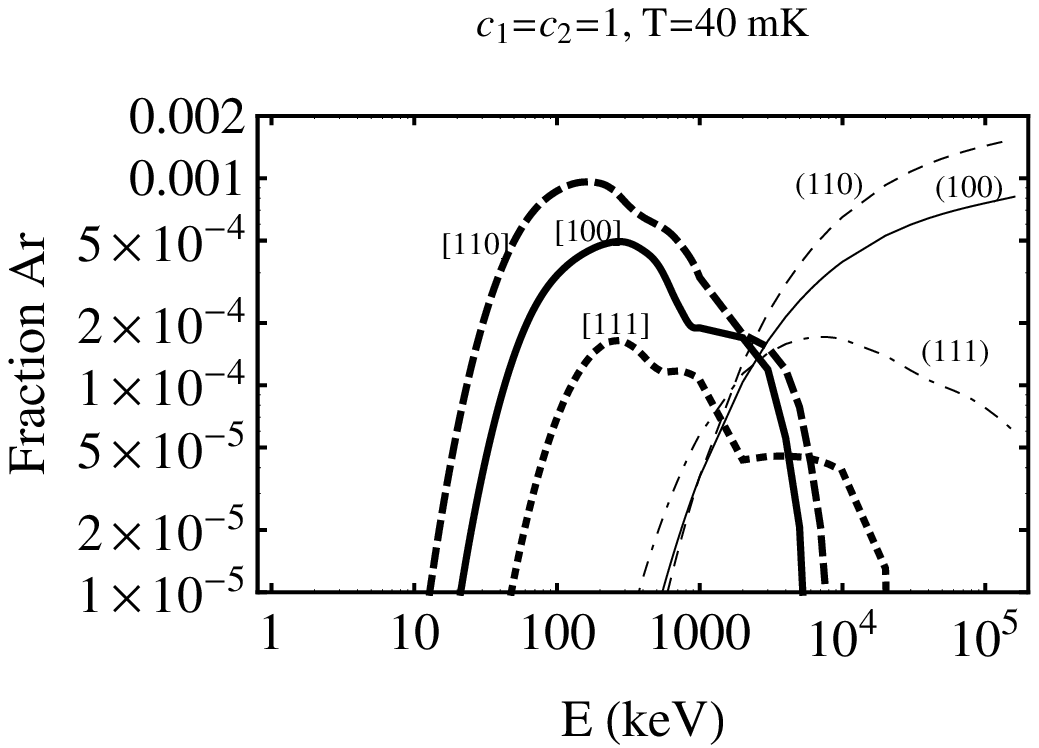}\\
  \includegraphics[height=145pt]{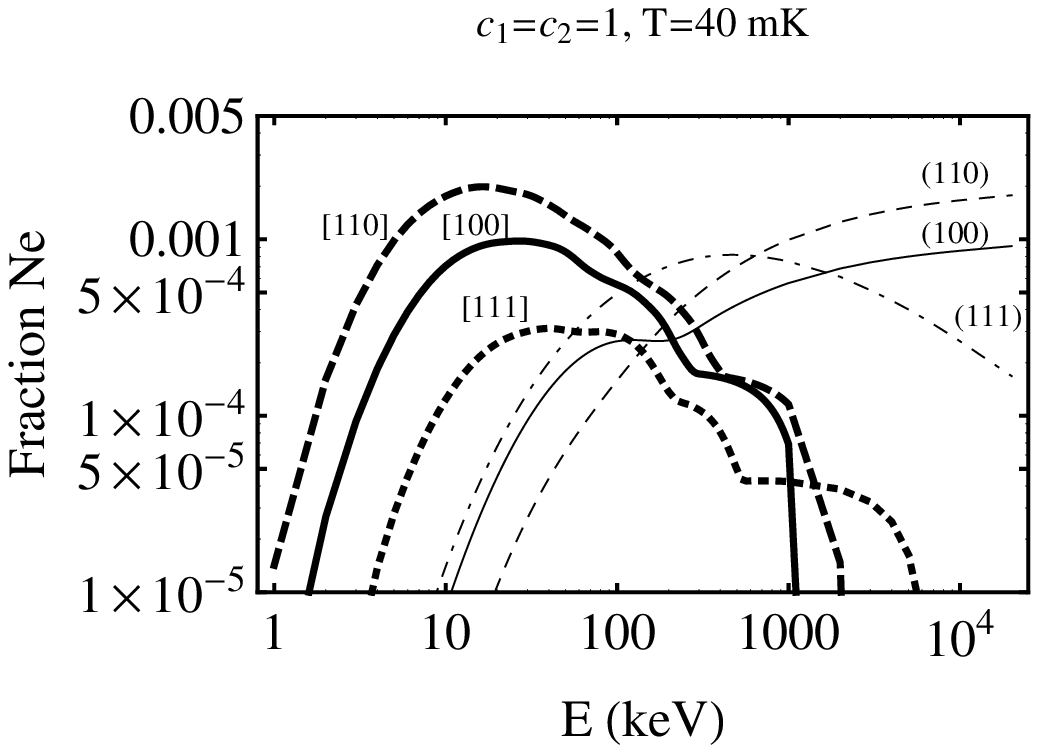}\\
  \caption{Channeling fraction of (a) Xe recoils at $T=77.2$ K, (b) Ar recoils at $T=40$ mK, and (c) Ne recoils at $T=40$ mK for single planar and axial channels, as a function of the recoil energy $E$ in the approximation of $c_1=c_2=1$.}
  \label{OneChannel}
\end{center}
\end{figure}

Fig.~\ref{OneChannel} shows the channeling fractions for several individual channels of (a) Xe ions propagating in a Xe crystal at $T=77.2$ K, (b) Ar ions propagating in an Ar crystal at $T=40$ mK, and (c) Ne ions propagating in a Ne crystal at $T=40$ mK (again with $c_1 = c_2 = 1$). The curves correspond to single axial or planar channels. Notice that at low energies channeling is dominated by axial channels, and at higher energies planar channels dominate.
\begin{figure}[t]
\begin{center}
  \includegraphics[height=145pt]{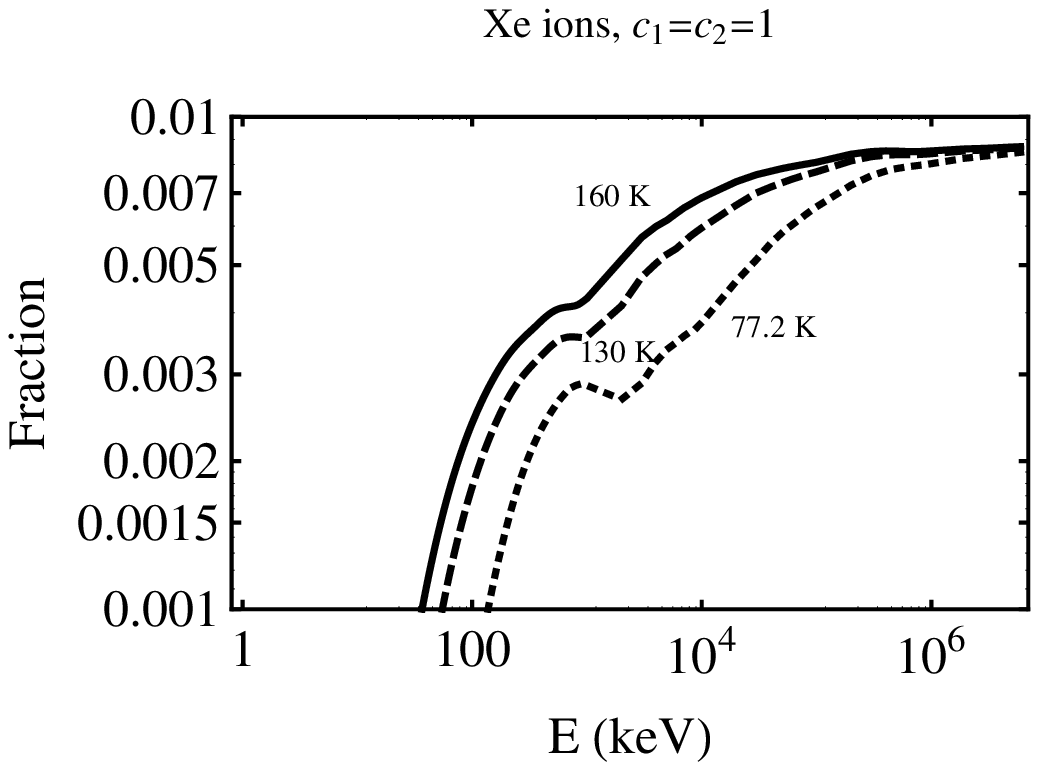}
  \includegraphics[height=145pt]{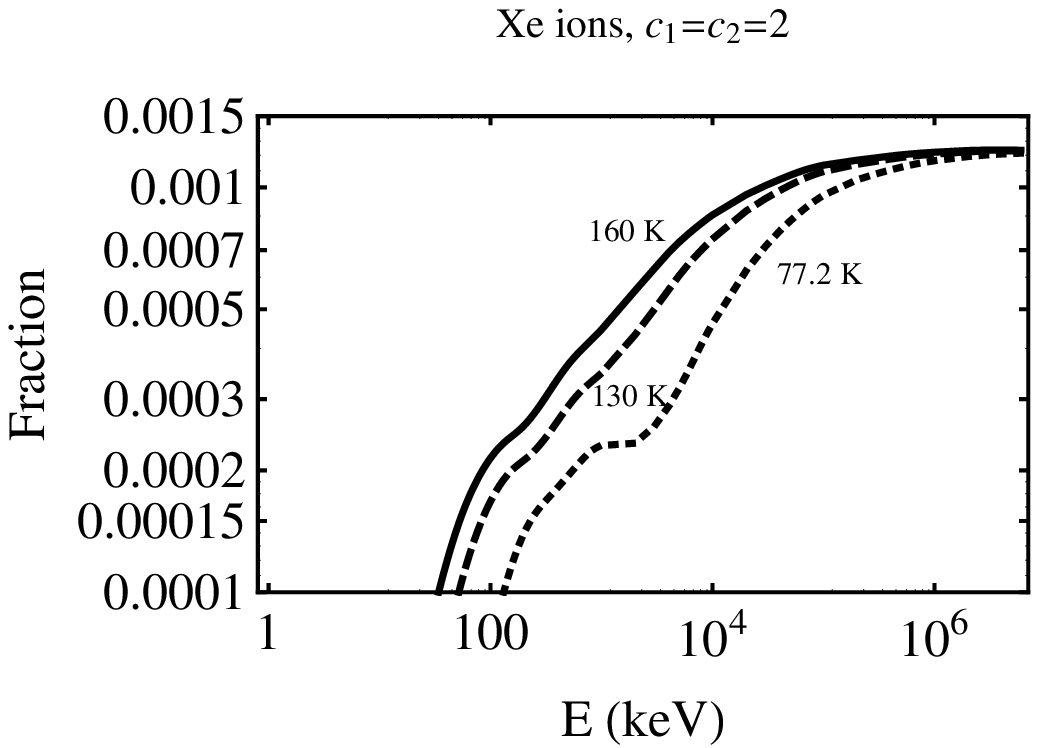}\\
  \includegraphics[height=145pt]{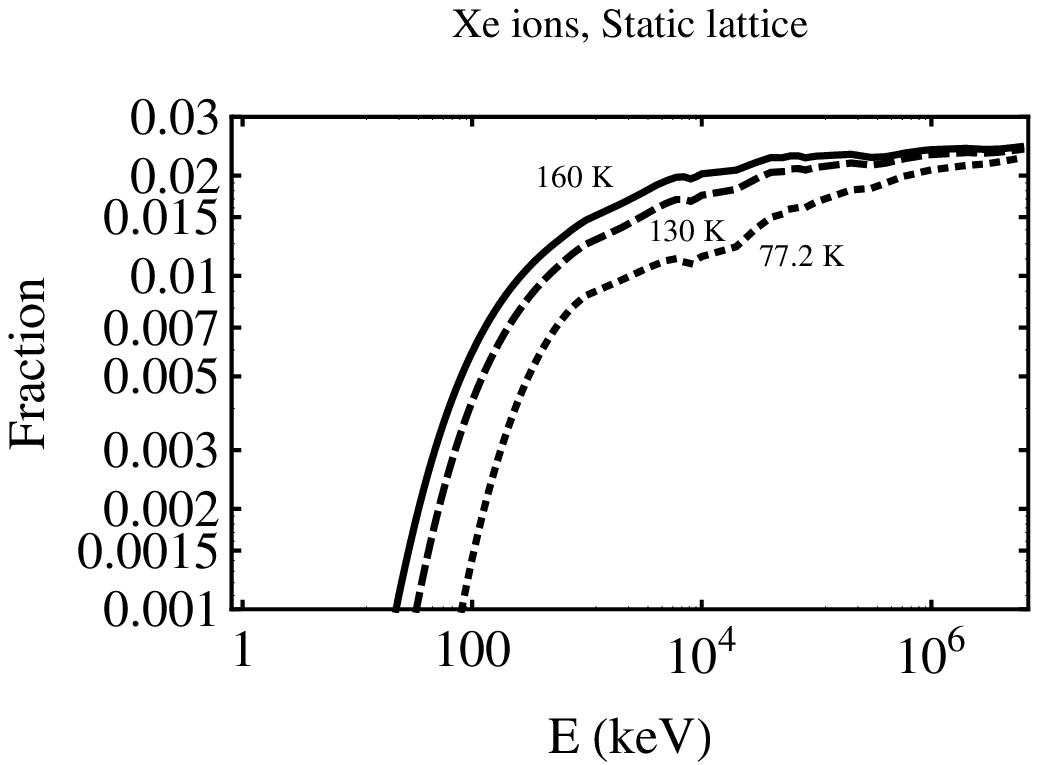}\\
  \caption{Channeling fraction of Xe recoils as a function of the recoil energy $E$ for $T=160$ K (solid line), 130 K (dashed line), and 77.2 K (dotted line) in the approximation of (a) $c_1=c_2=1$, (b) $c_1=c_2=2$ and (c) static lattice with $c_1=c_2=0$.}
  \label{FracXe-DiffT}
\end{center}
\end{figure}
\begin{figure}[h]
\begin{center}
  \includegraphics[height=145pt]{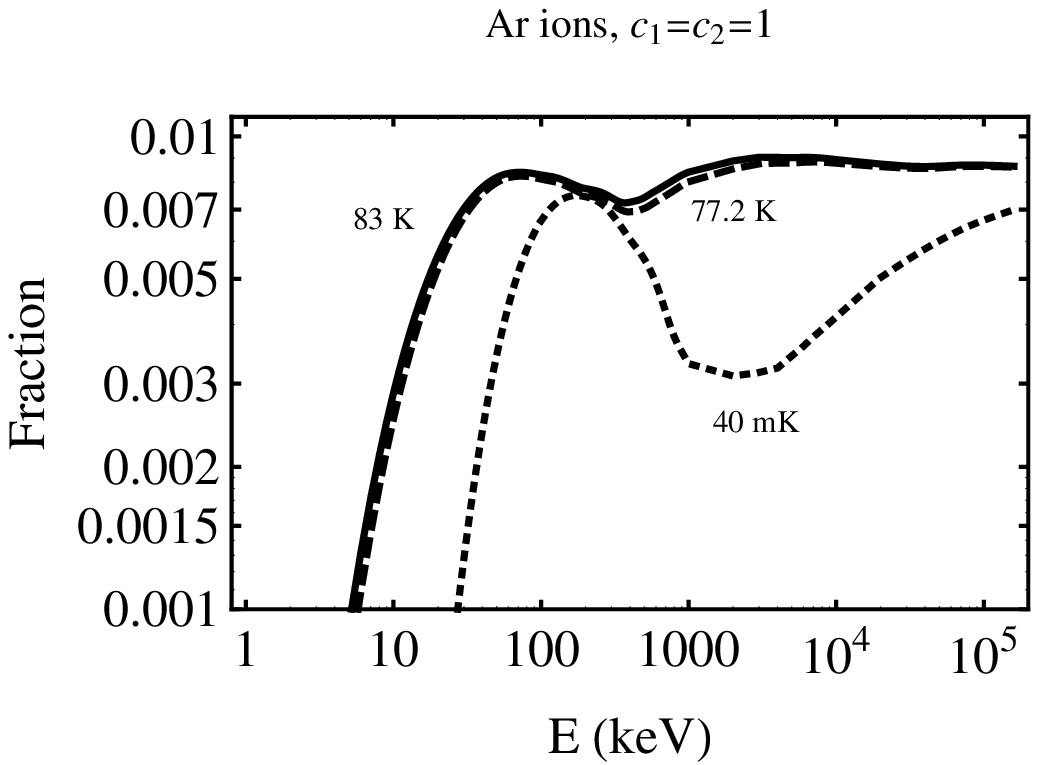}
  \includegraphics[height=145pt]{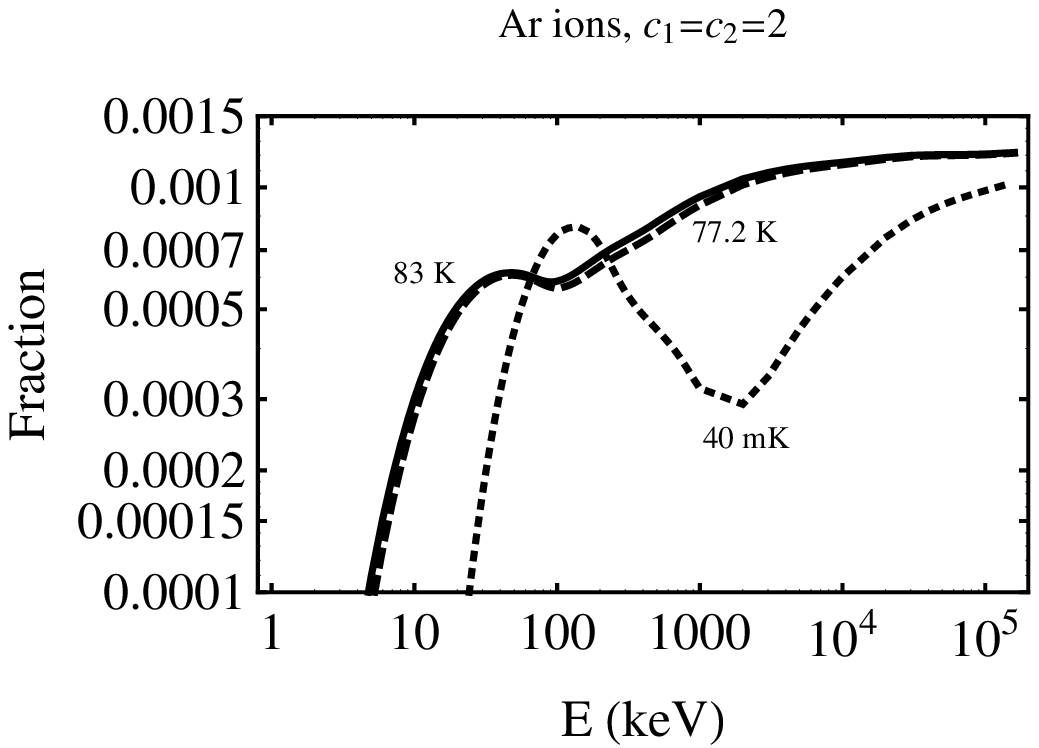}\\
  \includegraphics[height=145pt]{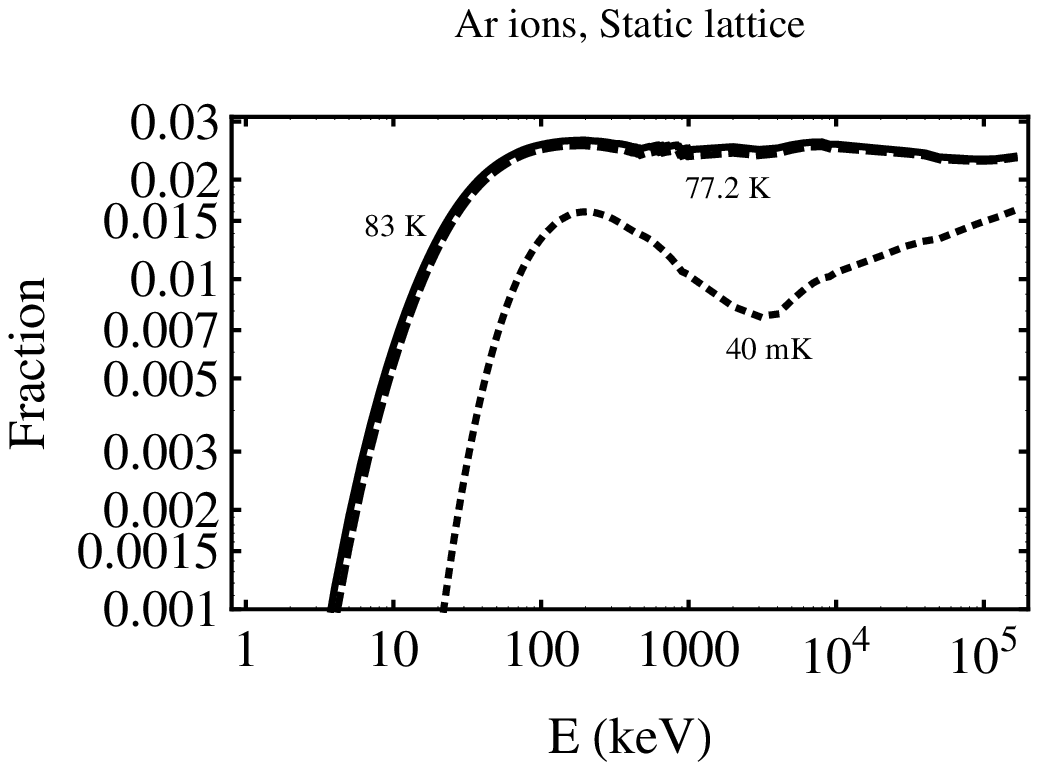}\\
  \caption{Channeling fraction of Ar recoils as a function of the recoil energy $E$ for $T=83$ K (solid line), 77.2 K (dashed line), and 40 mK (dotted line) in the approximation of (a) $c_1=c_2=1$, (b) $c_1=c_2=2$ and (c) static lattice with $c_1=c_2=0$.}
  \label{FracAr-DiffT}
\end{center}
\end{figure}
\begin{figure}[h]
\begin{center}
  \includegraphics[height=145pt]{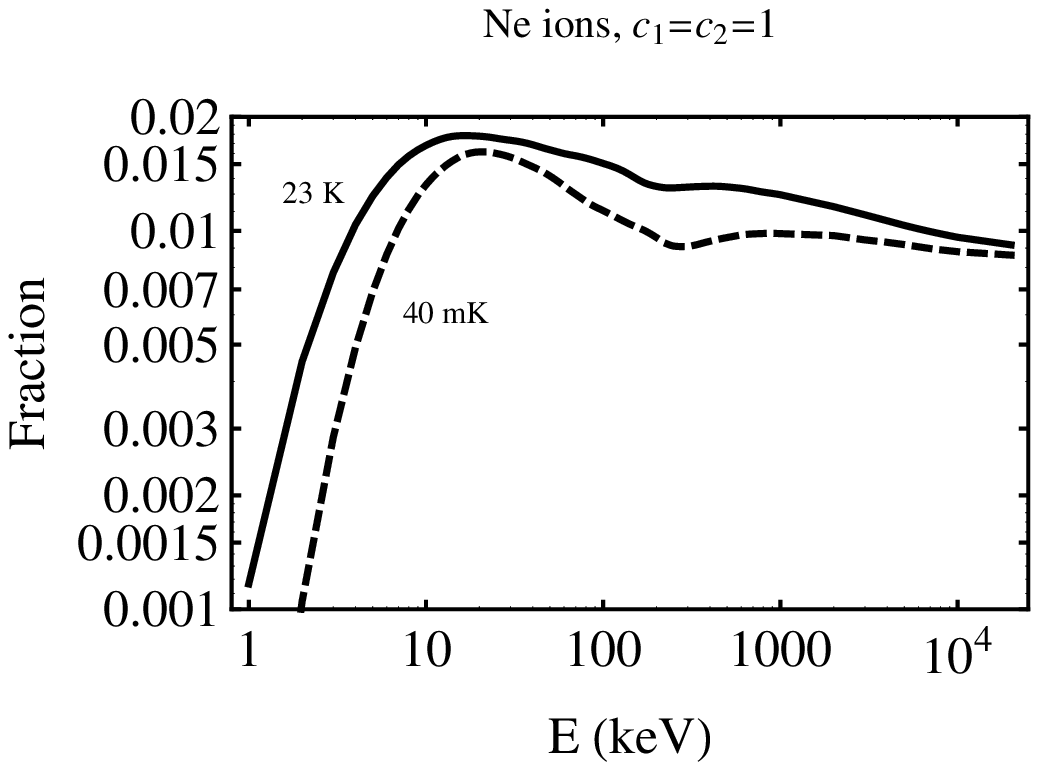}
  \includegraphics[height=145pt]{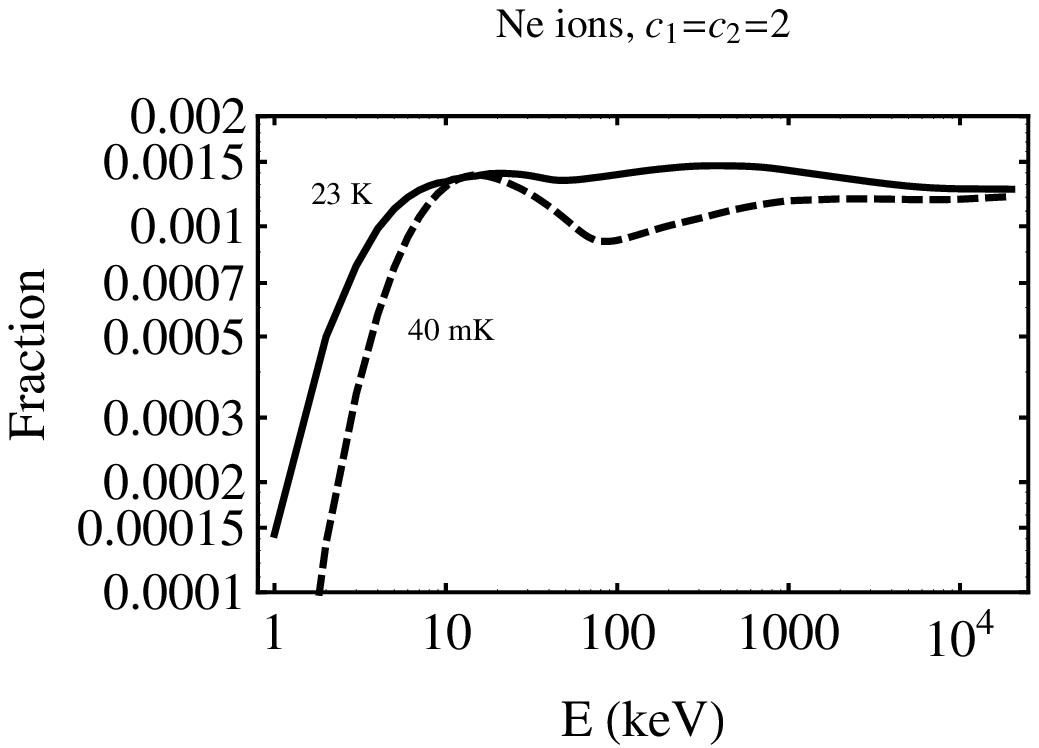}\\
  \includegraphics[height=145pt]{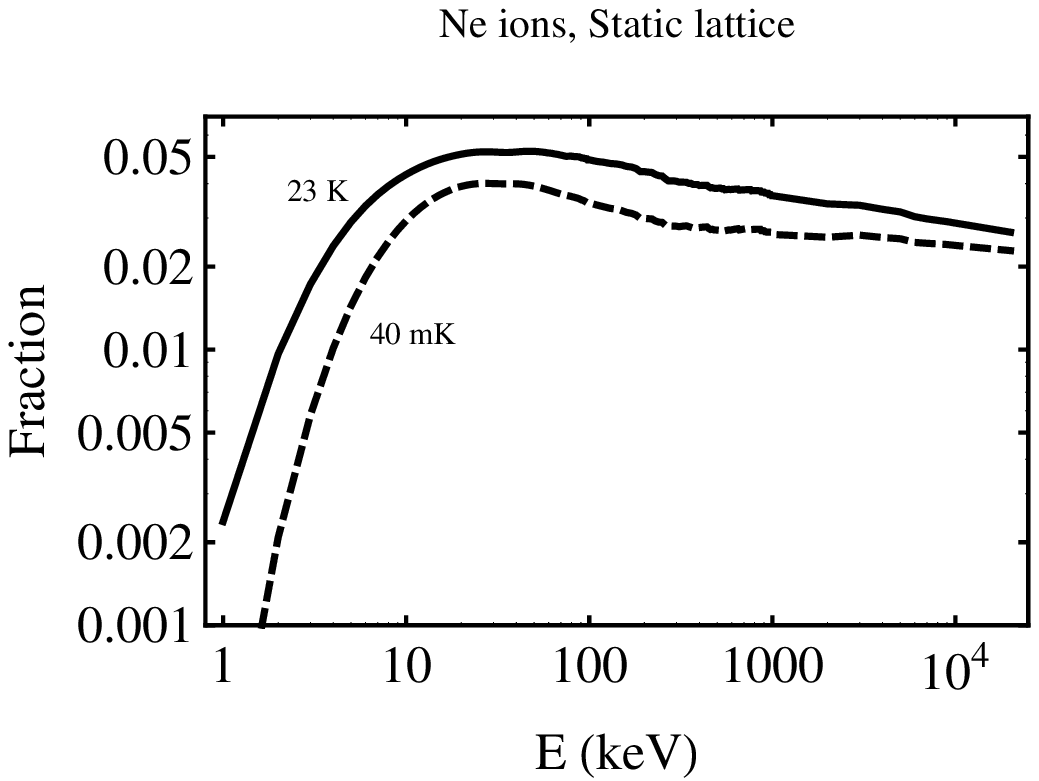}\\
  \caption{Channeling fraction of Ne recoils as a function of the recoil energy $E$ for $T=23$ K (solid line) and 40 mK (dashed line) in the approximation of (a) $c_1=c_2=1$, (b) $c_1=c_2=2$ and (c) static lattice with $c_1=c_2=0$.}
  \label{FracNe-DiffT}
\end{center}
\end{figure}

The geometric total channeling fractions of Xe, Ar, and Ne ions as  function of the  recoil energy are shown in Figs.~\ref{FracXe-DiffT}, \ref{FracAr-DiffT}, and \ref{FracNe-DiffT} respectively. For each crystal, we include three possibilities for  thermal effects: (a) $c_1=c_2=1$, a reasonable middle ground, (b) maximum effects, i.e. $c_1=c_2=2$, and (c) $c_1=c_2=0$, which corresponds to the unrealistic case of not having thermal effects in the lattice.

Notice that we have not considered the possibility of dechanneling of initially channeled ions due to imperfections in the crystal. Any mechanism of dechanneling will decrease the fractions obtained here.

As we see in Fig.~\ref{FracXe-DiffT} to \ref{FracNe-DiffT} the channeling fraction increases with energy, reaches a maximum at a certain energy, then has a dip and finally raises again. The maximum reflects the shape of the single channeling fractions, which all have maxima. These happen because the critical distances decrease with the ion energy $E$, which makes channeling more probable, while the critical angles also decrease with $E$, which makes channeling less probable. At low $E$ the critical distance effect dominates, and at large $E$ the critical angle effect dominates. At the maximum of the channeling fraction the axial channels dominate, in particular the channels $[110]$ and $[100]$ (as seen in Fig~\ref{OneChannel}).  The dip and the raise result from having multiple axial and planar channels contributing to the channeling fraction. At lower $E$ axial channels dominate and at higher $E$ planar channels dominate. The dip happens at the cross-over between both types of channels, when as the energy increases the contribution of axial channels dies out and that of planar channels is increasing. This increase causes the subsequent raise in the channeling fraction as the energy increases further.

 As shown in Figs.~\ref{FracXe-DiffT} (a), \ref{FracAr-DiffT} (a), and \ref{FracNe-DiffT} (a), the channeling fractions are never larger than 1\% for Xe and Ar and never larger than 2\% for Ne (with $c_1 = c_2 = 1$).

Temperature effects computed  with $c_1 = c_2 = 1$ are in the low range of what is found in other materials and propagating ions. Considering larger values of  $c_1$ and $c_2$, the channeling fractions are smaller. Figs.~\ref{FracXe-DiffT} (b), \ref{FracAr-DiffT} (b), and \ref{FracNe-DiffT} (b) show that with $c_1=c_2=2$, the maximum channeling fraction for Xe ions at 160 K, Ar ions at 83 K, and Ne ions at 23 K would be below 0.2\%.  However, since we do not know which are the correct values of the crucial parameters $c_1$ and $c_2$ for Xe, Ar, and Ne, we could ask ourselves how the channeling fractions would change if the values of these parameters would be smaller than 1 (although for other materials and propagating ions the values of these parameters were found to be between 1 and 2). The values of $c_1$ and $c_2$ cannot be smaller than zero, thus  Figs.~\ref{FracXe-DiffT} (c), \ref{FracAr-DiffT} (c), and \ref{FracNe-DiffT} (c) show the upper bounds on the geometric channeling fraction, obtained by setting $c_1=c_2=0$, namely by neglecting thermal vibrations of the lattice (which make the channeling fractions smaller as $T$ increases) but including the thermal vibrations of the nucleus that is going to recoil (which make the channeling fraction larger as $T$ increases). Although it is physically inconsistent to take only the temperature effects on the initial position but not on the lattice,   we do it here because using $c_1=c_2=0$, namely a static lattice, provides an upper bound on the channeling probability with respect to that obtained using any other non-zero value of $c_1$ or $c_2$. Even in this case, the channeling fractions cannot be larger than 5\%.

The critical angles depend on the temperature through the minimum distances of approach: as these increase with increasing temperatures, the critical angles decrease, what makes the channeling fraction smaller. However, there is a second temperature effect which makes the channeling fractions larger as the temperature increases: the vibrations of the atom which collides with the dark matter particle. Depending on which of the two competing effects is dominant, the channeling fraction may either increase or decrease as the temperature increases. Increasing the temperature of a crystal usually increases the fraction of channeled recoiling ions, but when the values of $c_1$ and $c_2$ are large (i.e. close to 2) so the critical distances increase rapidly with the temperature, the opposite may happen (see Fig.~\ref{FracAr-DiffT} (b)).

 To conclude, let us remark that the analytical  approach used here can successfully describe qualitative features of the channeling and blocking effects, but should be complemented by data fitting of parameters and by simulations to obtain a good quantitative description too.  Thus our results should in the last instance be checked by using some of the many sophisticated Monte Carlo simulation programs implementing the binary collision approach or mixed approaches.

 \section*{Acknowledgments}
N.B. and G.G. were supported in part by the US Department of Energy Grant
DE-FG03-91ER40662, Task C.  P.G. was  supported  in part by  the NFS
grant PHY-0756962 at the University of Utah. We would like to thank Dr.~Jonghee~Yoo for drawing our attention to the use of solid xenon in dark matter detection and for providing important information on the Solid Xe R\&D Project.

\appendix
\addappheadtotoc

\section{Crystal structure and other data for Xe, Ar, and Ne}

 Solid Xe, Ar and Ne have a face-centered cubic (f.c.c.) lattice structure with 4 atoms per unit cell. The lattice constant of Xe, Ar and Ne crystals are $a_{\rm lat}^{\rm Xe}=0.620$ nm at $T=75$ K~\cite{Sears:1962}, $a_{\rm lat}^{\rm Ar}= 0.525$ nm and $a_{\rm lat}^{\rm Ne}= 0.442$ nm at $T=4.2$ K and atmospheric pressure~\cite{Henshaw:1958}.

The atomic mass and atomic numbers of Xe, Ar, and Ne are $M_{\rm Xe}=131.29$ amu, $M_{\rm Ar}=39.948$ amu, $M_{\rm Ne}=20.1797$ amu, $Z_{\rm Xe}=54$, $Z_{\rm Ar}=18$ and $Z_{\rm Ne}=10$.

The Thomas-Fermi screening distance for an ion expelled from a lattice site in the crystal scattering on another atom in the same crystal is $a_{\rm XeXe}=0.4685 (Z_{\rm Xe}^{1/2} + Z_{\rm Xe}^{1/2})^{-2/3}=0.007808$ nm for Xe, $a_{\rm ArAr}=0.4685 (Z_{\rm Ar}^{1/2} + Z_{\rm Ar}^{1/2})^{-2/3}=0.01126$ nm for Ar  and $a_{\rm NeNe}=0.4685 (Z_{\rm Ne}^{1/2} + Z_{\rm Ne}^{1/2})^{-2/3}=0.01370$ nm for Ne.

To compute the interatomic spacing $d$ in axial directions and the interplanar spacing $d_{\rm pch}$ (``pch'' stands
for ``planar channel'') in planar directions, we have to multiply the lattice constant by the following coefficients~\cite{Gemmell:1974ub}:
\begin{itemize}
  \item Axis: $<$100$>: 1$ , $<$110$>: 1/\sqrt{2}$ , $<$111$>: \sqrt{3}$
  \item Plane: $\{100\}: 1/2$ , $\{110\}: 1/2\sqrt{2}$ , $\{111\}: 1/\sqrt{3}$
\end{itemize}

 The Debye temperatures  of Xe, Ar and Ne are $\Theta_{{\rm Xe}}=55$ K, $\Theta_{{\rm Ar}}=85$ K and $\Theta_{{\rm Ne}}=63$ K~\cite{DebyeXe, DebyeAr}, and the crystals in the Solid Xe R\&D Project experiment will be operating at a temperature of 77.2 K or higher~\cite{Yoo-Private}.

\end{document}